# Designing an ELearning Portal for Developing Countries: An Action Design Approach

**Jay Douglas**
School of Engineering and Information Technology,
Australian Defence Forces Academy,
University of New South Wales
Campbell ACT 2600, Australia
Email: jay.douglas@defence.gov.au

**Dr Ahmed Imran**
School of Engineering and Information Technology,
Australian Defence Forces Academy,
University of New South Wales
Campbell ACT 2600, Australia
Email: a.imran@adfa.edu.au

**Dr Tim Turner**
School of Engineering and Information Technology,
Australian Defence Forces Academy,
University of New South Wales
Campbell ACT 2600, Australia
Email: t.turner@adfa.edu.au

## Abstract

This paper presents the first phase of a study on using current eLearning trends in course design to overcome challenges in eLearning within developing countries, particularly for non-tertiary training providers. The paper outlines the research and development of an ICT artefact using the Action Design Research method. The artefact will later be deployed and evaluated. The paper discusses the previous research that has categorised challenges in eLearning in developing countries and explores how these can be overcome through course and element design. Course design includes training development, platform selection and platform hosting, particularly in the context of developing countries. Element design looks at the creation of content that should be available, practical and viable to both the user and developer through the exploitation of current eLearning trends. The paper concludes with the broad design of the ICT artefact that will be implemented for phase two of the study.

## 1     Introduction

The use of eLearning to deliver a wide variety of training and education, via a range of mediums, to a diverse audience in a flexible, economical and collaborative manner is now a mainstream option for education delivery throughout the developed world (Maldonado et al. 2011; Quimno et al., 2013). With advances in technology, connectivity and mobile computing, the digital divide is not what it once was (United Nations 2012; Wijetunga 2014) making the virtual learning environment more available through a rise in the provision of open source software and complimentary hosting options. ELearning is looking increasingly attractive to developing countries (Maldonado et al. 2011), but there are still challenges, such as ICT use, cultural differences to western theories of andragogy, and adoption of new training methods (Quimno et al., 2013). With its prevalence in developed countries, eLearning developers have focused on technical and functional characteristics. In developing countries, eLearning developers must understand how these technological and functional tools can be contextualised to meet their needs.

It is important to understand the needs and background of the learner prior to developing training. This is no different for eLearning and more so in developing countries, largely because of the nature of the society to which they belong (Hofstede 2001). The complexity around the socio-cultural context and the technological environment challenges eLearning implementers to step away from the one size fits all approach. However, it has been shown that with the decrease in the digital divide (United Nations 2012; Wijetunga 2014) and the rise in the use of mobile computing and social networking applications, citizens of developing countries are increasingly connected and in a position to benefit from eLearning. This increased connectivity raises the question: can the challenges of eLearning in developing countries be overcome with course design, using current eLearning trends?

The study presented here shows the use of course design incorporating current eLearning trends to overcome the challenges to eLearning in developing countries. The study will also help to provide a better understanding of eLearning design for developing countries.

The study is being conducted over two phases using the Action Design Research (ADR) method. The first phase, presented in this paper, is the research and development of an ICT artefact designed for eLearning in developing countries. The paper presents the initial research conducted regarding the use of eLearning in developing countries to determine the challenges and possible solutions. The paper then outlines the ADR method and the background behind the Bangladesh case study. The paper then explores the development of an ICT artefact, specifically the concept of course design to meet the challenges of technology and course development, and element design to meet the challenges of context and the individual. The second phase of the project will deploy and evaluate the ICT artefact.

## 2 Initial Research

### 2.1 ELearning and its Evolution

The definition of eLearning varies between sources (Andersson and Grönlund 2009; Clark and Mayer 2011; Dašić et al. 2010; Driscoll 2002) but the general theme is that eLearning is learning through instructional systems that are designed using andragogy and technology, to impart knowledge through the use of digital devices. Early examples of eLearning included systems as simple as CD-ROMs delivered in the mail with corresponding texts. Through ongoing development of andragogy and technology, the term eLearning now implies a more complex delivery method (Moore et al. 2011). ELearning is no longer unidirectional content delivery, but is interactive, engaging and easy to use products delivered on digital devices. Students view the learning material, and interact with it, other users, and the instructional system itself. The trends in the use of eLearning now establish, at a minimum, an application or platform that provides multimedia content relevant to the learning objectives, delivered using instructional methods that are either instructor-led, self-paced or a combination of both, and through a digital device from a local drive, local server, or the Internet (Clark and Mayer 2011).

Groups like mooc.org, edX, moodle.com, and Google are working to support eLearning trends of more readily available eLearning facilities for course designers and students, through the use of open source platforms and hosting opportunities. Other current eLearning trends such as gamification and m-learning are starting to increase, taking advantage of the popular social media use in developed countries. These eLearning trends work by making learning motivating, engaging and accessible everywhere (Garcia-Cabot et al. 2015; Hamari et al. 2014). These latest eLearning trends require more technical expertise in design and development, but ready to use internet-based applications such as social media and video hosting can be viable alternatives, until more advanced gamification and m-learning design tools become more readily available.

### 2.2 ELearning for Developing Countries

ELearning has the potential to make a positive contribution to learning in the developing world by increasing the access to education, particularly for marginalised groups in rural or isolated areas, in spite of shortages in teachers and facilities (Andersson and Grönlund 2009).

*"eLearning has the* potential *to reach out to more people than hitherto possible with conventional learning methods, not only geographically, but demographically. eLearning is a valuable development tool that can reach out to the underprivileged and help build a culture of life-long learning into society at large" (*Rattakul and Morse 2005, p 342).

Studies have shown that some universities in developing countries have been early adopters of eLearning that has proven beneficial for students (Bhuasiri et al. 2012; Pandey 2013; Quimno et al., 2013). However, the vocational and workplace training establishments (non-tertiary training providers) in developing countries have not yet been sufficiently explored. In Australia, Registered Training Organisations (RTOs) range from public vocational training schools to private companies training their staff and even the government agencies such as the Australian Defence Force. These RTOs all use eLearning to deliver a variety of training to a diverse audience spread around the world. This adoption of eLearning is not as widely prevalent in similar organisations in developing countries. This research seeks to address that gap by exploring why and how eLearning can be implemented in non-tertiary training providers in developing countries. The case study for this research is an example of the non-tertiary training providers, the Bangladesh Public Administration Training Centre (BPATC) that provides the eGovernment Management Course (EGMC) as one of a range of courses for the Bangladesh Public Service.

## 2.3  Challenges for Developing Countries

There is a broad range of research conducted on the area of eLearning in developing countries. Andersson and Grönlund (2009) undertook a thorough review of academic papers to categorise the challenges of eLearning and the differences between developing and developed countries. Four categories grouped the challenges: individual, course, contextual and technological. By comparing the numbers of papers that research each challenge, it is seen that *context* and *technology* are more widely researched in developing countries (Figure 2.1). This focus could indicate that the challenges relating to technology and context may be greater for developing countries than their developed counterparts, whilst the challenges relating to the course are similar and to the individual is significantly less in developing countries.

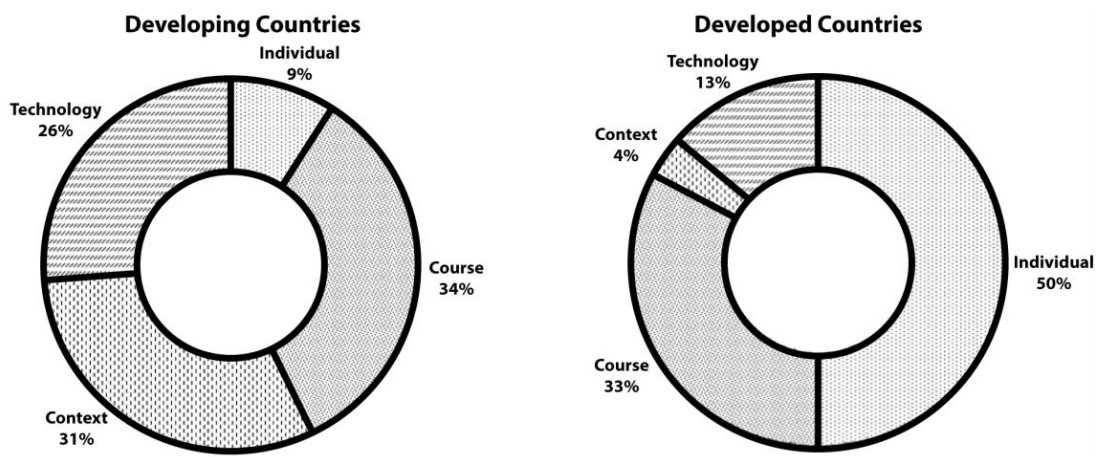

*Figure 2.1: ELearning research focus developed versus developing countries (based on Andersson and Grönlund 2009)*

Due to the larger interest in technology and context challenges, these areas will be explored in greater detail, whilst understanding that the course and individual challenges still play a part for developing countries.

Sife et al. (2007), Andersson and Grönlund (2009), Islam and Selim (2006), and Ssekakubo et al. (2011) have shown that some of the common challenges when using technology in developing countries are the lack of a systematic approach to the implementation of ICT, awareness and attitudes when introducing ICT, administrative support, technical support, staff development and inadequate funds. Many of these studies were conducted over five years ago. Given the advances in technology in recent times, especially the speed and spread of technology reducing the digital divide (Gemmell et al. 2015; United Nations 2012; Wijetunga 2014), it is likely that these concerns may be overcome. Information technology trends and approaches that might now apply to delivering eLearning in developing countries include the increase in mobile computing and connectivity, the availability and competitiveness of cloud computing providers, freely-available open source applications, and the changing attitudes of government and the commercial sector toward ICT use. For example, some developing countries, such as Tanzania and Bangladesh have used strategic level initiatives, such as incentives and tax breaks for ICT infrastructure and equipment, to build their ICT capacity as well as whole of government change toward ICT use in its provision of services (Sife et al. 2007; Imran et al., 2013b). The decreasing costs of tablet and laptops computers is making it easier for individuals to enter the market (Romiszowski 2013). The use of mobile devices, tablets and laptops that have their own power source are proving to be a work around for poor electrical and hard-wired communications infrastructure typical of developing countries (Andersson and Grönlund 2009).

Greater access to mobile computing and wireless connectivity has led to a greater uptake of technology (Williams 2013). The increasing presence of the digital native (Romiszowski 2013) or digital citizen (Imran et al., 2013b) is proving that global citizens from both developing or developed countries are increasing their use and skills in ICT and are expecting their commercial and government service providers to do the same. These factors show how the digital divide, not only on an individual level, but at a national level, is now beginning to close (United Nations 2012; Wijetunga 2014). This reduction in the digital divide means that there is an expanding number of users able to access virtual learning environments. The increasing uptake and changing attitudes toward ICT are expected to provide an increase in acceptance of eLearning as a viable alternate form of training for developing countries.

## 3  Research Method

The study presented here uses the ADR method (Sein et al. 2011) to research and design an ICT artefact, which will then be deployed and its findings evaluated. This paper represents the first phase, which explores the design of the eLearning tool for the EGMC as the case study for this research. This paper looks at research that has already been conducted on the challenges of eLearning in developing countries, focusing on the areas of andragogy, culture and ICT, as well as exploring the use of emerging and popular technologies, such as the use of social media and cloud computing. That research guides the design of the eLearning course development in the context of developing countries. The outcome of the first phase is the conceptual design of a prototype eLearning tool and ICT artefact, which will be used to deliver the first module of the EGMC. The second phase will be the development and deployment of the ICT artefact, and its evaluation. The study will use the inherent analytics provided in the eLearning tool and an embedded user survey to capture critical feedback and data from the students and facilitators on whether the designed use of specific eLearning concepts and content were successful in delivering the desired outcome.

### 3.1  The Case Study

The EGMC offered by the BPATC has been selected as the BPATC is a non-tertiary training provider and the EGMC has common goals with our research, being the improvement of developing countries through greater access to improved services through ICT. The overarching goal of the EGMC is to improve efficiency, productivity and the governance in the public sector of recipient countries through capacity building to implement effective e-government. A research project funded by the Australian Government aid agency, AusAID, identified the fundamental problem of a lack of knowledge of the strategic use of ICT systems for government business processes in Bangladesh. The lack of knowledge was a major barrier to e-government adoption and was seen as underpinning a range of other barriers such as poor infrastructure, low socio-economic conditions, and a lack of leadership (Imran and Gregor 2010). The second phase of the project aimed to institutionalise the knowledge transfer through effective curriculum and training for all Bangladesh public servants in collaboration with the BPATC.

The EGMC is the embodiment of that knowledge transfer through a traditional face-to-face classroom-based training course. The EGMC is a complete postgraduate-level course representing a semester-long course of instruction: that is, 13 weeks of 3 hours of lectures, tutorials, and workshops. It is supported by a new textbook on e-government management, the first of its kind, developed as part of the project. The course is offered to BPATC students one or more times a year, on the BPATC campus, delivered by trained BPATC teachers and supplemented by guest lecturers. The course relies upon the textbook and prepared course material to define and deliver the scope of interest of the course. Written assignments and in-class tests are used to assess the students' performance in the course.

The present research is part of the effort to elevate that successful training offering to the next level, converting the existing course to an eLearning platform (Imran et al., 2013a). Moving online will allow a much broader impact, not limited within Bangladesh. The online course intends to promote knowledge transfer to a large target audience, using culturally-tailored online education, which will be sustainable, repeatable and offer greater flexibility than the traditional classroom approach currently used.

### 3.2  The Action Design Research Method

The ADR method has been chosen as this research will centre on the development and evaluation of an ICT artefact. As outlined by Sein et al. (2011), the ADR will deal with two research challenges: identifying a problem, and constructing an artefact to address the problem. ADR was selected to facilitate a rigorous approach to designing an eLearning solution drawing on the insights of established research and applying them in a novel circumstance. In particular, the design approach allowed thoughtful selection of eLearning trends in developed countries in the creation of a specific solution to the researched problem domain. ADR uses four stages that are *Problem Formulation, Building, Intervention* and *Evaluation*, as well as continual *Reflection and Learning* and concluding with the *Formalisation of Learning*. We outline below how the ADR stages have been implemented in the reported research.

### 3.2.1 ADR Stages

**Problem Formulation.** As detailed thus far, there are specific challenges to using eLearning in the target circumstances, specifically, in non-tertiary training providers in developing countries. We seek to find design principles that are influential in eLearning course design for a target audience in developing countries. While there are conventional approaches to delivering eLearning available, these approaches are not tailored to meet the constraints present in developing countries. The problem being identified is whether there is a model for developing and delivering eLearning courses using these conventional tools that specifically addresses developing country constraints like infrastructure capacity, cultural and educational conventions around learning, and language and access limitations. There are two measures of success for this problem: effectiveness and efficiency. Effectiveness looks at the learning outcomes and if they were delivered, properly assessed, and assimilated. Other effectiveness measures include: the range of functions available to the facilitators, and the capacity of the technology to host the required content. Efficiency looks at the use of finances, resources, and time. Efficiency considers the costs associated with hardware, software and subscriptions, and the time involved, both in its general use and in the time it takes to train the learner and facilitator in its use. Efficiency also includes the speed and ease in which people are adapting and connecting through ICT via popular online applications.

**Building** will begin with the conceptual design of the EGMC using the design principles that have been discovered in this research and applying them to this unique context. The design will then guide the construction of the ICT artefact, which will be the functioning prototype that can be assessed for effectiveness and efficiency.

**Intervention** will be the deployment of the ICT artefact to participants. The students, facilitators and managers of the EGMC at BPATC will be invited to take part in the study. It is hoped that there will be a significant number of participants to provide valid results.

**Evaluation** will use the analytics provided by the ICT artefact, as well as the feedback provided by the students and facilitators, as the data source. Measures of effectiveness will be derived from de-identified user performance results and analysis of logs showing interaction by users engaging with the presented material (durations, repeat accesses, skipped material, etc). Effectiveness for learning developers will be derived from specific feedback and reflection during the design and development of the course. Measures of efficiency will be derived from records of expenses and logs of time by the developer as well as analysis of logs from the artefact showing connection sources, session times, and session drop-outs of the course in use.

**Reflection and Learning** will be conducted at the end of each cycle, not to prove there is a solution but to ensure continual improvement of the ICT artefact. Insights from that reflection will then lead into the planning for the next cycle.

**Formalisation of Learning** is the culmination of all the stages, cycles and the ADR as a whole. We will use the findings from all the cycles and reflect on these against the original problem formulation. We expect to determine the design principles that meet the requirements of effectiveness and efficiency in the context of developing countries. Reflection will involve a mix of quantitative and qualitative strategies, used concurrently while examining the course and survey data. Quantitative data will be gathered through direct closed questioning and the course platform analytics, whereas the qualitative data will be collected through open questioning of participants.

### 3.2.2 Study Cycles

The study will progress through the six ADR stages using cycles. These are:

- Cycle 1. Selection of the eLearning platform, delivery method and content elements.
- Cycle 2. ICT artefact design, which will be reviewed by the research supervisors for suitability.
- Cycle 3. Course content will then be written into the ICT artefact and reviewed for educational accuracy and user interface usability.
- Cycle 4. The complete ICT artefact will be deployed to Bangladesh for users to complete and provide feedback through questionnaires and interviews. This cycle will be repeated if there is available time and participants.
- Cycle 5. The evaluation of the data obtained through the survey and ICT artefact analytics.
- Cycle 6. The consolidation of all cycles to communicate the Formalisation of Learning.

Each of these cycles can be repeated as required, as shown in Figure 3.1. This paper reports on the Reflections and Learnings of the first two stages (first three cycles) of the ADR approach.

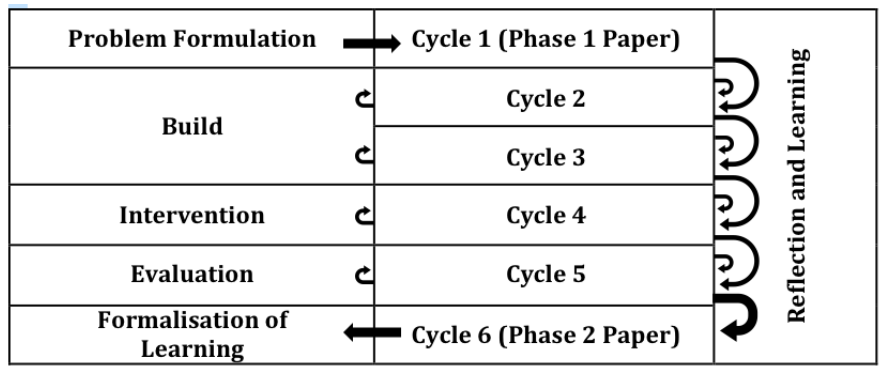

*Figure 3.1: Study Cycles within the ADR Stages (based on Sein et al. 2011)*

## 4　ICT Artefacts in eLearning

Cycle 1 of the ADR approach we have adopted requires the selection of the eLearning platform, delivery method, and content elements. A broad-brush one-size-fits-all approach cannot be applied to all developing countries and all possible courses within them. The EGMC case study will be used to explore the design principles that are influential in eLearning course design for developing countries. The eLearning design needs to be considered from two perspectives, the *course design* and *element design*. These two perspectives also include Andersson and Grönlund's (2009) categories, where course design encompasses technology and course challenges and element design includes context and individual challenges. Table 4.1 provides a list of decisions that are made in eLearning design, with a particular focus of the context of developing countries:

| Decisions | Options |
|---|---|
| Type of platform | Open source platform |
|  | Custom designed TDP |
|  | COTS product |
|  | Manual course management |
| Platform delivery | Local drive |
|  | Local server |
|  | Internet |
| Content elements | Text on screen |
|  | Video/animation |
|  | - Piece to camera |
|  | - Kinetic typography |
|  | - Info-graphics |

|  | Forums |
|  | Collaborative workspaces |
|  | Activities |

*Table 4.1. Design decisions*

## 4.1 Course Design

Cycle 2 of the ADR approach requires the development of a course design. Course design is made up of three areas: training development, platform selection, and platform delivery. These areas represent the course and technology challenges faced by developing countries. They are not considered in isolation. Training development focuses on the course challenges whilst still keeping in mind the technology challenges. Platform selection looks at the course and technology challenges together. Platform delivery focuses on the technology challenges, with course challenges still in mind.

### 4.1.1 Training Development

In this research, the training need has already been identified and is embodied in the traditional face-to-face courses already running. Therefore, the training development can begin immediately. In other instances, training development must be preceded by training needs analysis (Borotis and Poulymenakou 2004). The challenge for this research is the re-development of traditional course content into the eLearning format. To build an effective and efficient eLearning package, the development will be different to that for traditional instruction (Andersson and Grönlund 2009). There is a sufficient range of content available for course development from the existing course. The options available for eLearning content elements—such as spoken or printed text, images, videos, activities, guided research, social media, webinars and so on—are constrained by the capability of the course developer, manager, facilitator and user to produce or consume the content in those forms. Similarly, the selection of an eLearning platform is limited by connection speeds and the availability of staff with familiarity in the platform. The platform's capacities might also constrain the selection of content elements. The promotion and demonstration of contextual eLearning design principles for developing countries is necessary to give the facilitators and managers the knowledge of what is available and how to transition from traditional curriculum to the eLearning environment (Khan et al. 2012). To overcome the course challenges, the course designer must know what technologies are available, practical and viable, and they must understand training development for the transition into the eLearning field.

### 4.1.2 Platform Selection

Platform selection forms an integral part of deciding how to deliver the curriculum through the use of ICT. The course designer makes decisions on the type of platform required informed by an understanding of what needs to be delivered, to how many people, in which locations, and the training management data that needs to be captured. Platform selection balances technological abilities and limitations with a fit-for-purpose course design. ELearning delivery involves planning, implementing, facilitating, assessing, and monitoring student learning. ELearning models that attain support those objectives are known by a number of different names, such as: LMS (Learning Management System), CMS (Content Management Systems), LCMS (Learning Content Management System), VLE (Virtual Learning Environment), VLS (Virtual Learning System), Learning Portals, and so on (Kapp 2003). The EGMC case study will explore the use of some common implementations of eLearning delivery platforms, including: open source LMS, MOOC (Massive Open Online Course), COTS (Commercial off the Shelf) products, bespoke TDPs (Training Delivery Platforms), and Instructor Managed eLearning. Table 4.2 presents our summary of the advantages and disadvantages of these training delivery methods in the context of developing countries.

| Platform | Advantages | Disadvantages |
| --- | --- | --- |
| Open Source LMS | Widely available | Can be difficult to make an open course |
|  | Open source | Requires some technical setup |
|  | Customisable | Requires mid level facilitator |
|  | Easy to use | intervention |
|  | Wide variety of content | Course management intensive |
| MOOC | Easily obtained | Requires some technical setup |
|  | Open source | Content must be designed carefully to ensure active participation |
|  | Customisable |  |
|  | Easy to use | Discussions can be difficult |
|  | Wide variety of content | Grading can be difficult |
|  | Can easily enrol many students | Lack of interaction |
|  | Less facilitator intervention |  |
| COTS Product | Reliable | Expensive |
|  | Current due to competitive nature of the market | Cannot access underlying source code |
|  |  | Provider may discontinue product |
|  | Can often be hosted by the developer | May have restrictions on customisation |
| Bespoke TDP | Reliable | Expensive |
|  | Current due to competitive nature of the market | Cannot access underlying source code |
|  |  | May require ongoing developer maintenance |
|  | Can often be hosted by the developer |  |
|  | Customisable |  |
| Instructor Managed ELearning | Cheap | Facilitator intensive |
|  | Focus on the content | Requires smaller enrolment sizes |
|  | Customisable |  |

*Table 4.2. Pros and cons of specific platforms (from www.educause.edu; joedeegan.blogspot.com.au; adulated.about.com; blog.geteverwise.com)*

### 4.1.3 Platform Delivery

To ensure that the selected eLearning platform performs as required, a hosting method must be selected. The most common options for eLearning platform hosting are: a local server, on the Internet, or a hybrid of the two (https://seertechsolutions.com/our-services/hosting-options). The ICT challenges of developing countries, as explained earlier, often mean that there may not be sufficient infrastructure to host such a platform reliably or cost-effectively (Sife et al. 2007). Unreliability can affect the course facilitator and the participant and the reputation of the use of ICT in learning generally (Romiszowski 2013). There are many open source eLearning platforms that could be applied to this problem. Using these solutions implies some form of local support and infrastructure (i.e. servers) and the attendant issues of data security and reliability (Romiszowski 2013; Pocatilu et al. 2009). Alternatively, those eLearning platforms, and others, are available through online hosting; i.e. on the cloud. The trade-off involved in selecting local or cloud-based servers for hosting is accessibility. Local servers allow operation and connectivity to the eLearning course for facilitators and students locally (i.e. from on-campus) when Internet access is unavailable. Local server installations are risky because they rely on the local institution's infrastructure reliability and the availability of suitably trained technical staff. These matters are a considerable risk in developing country contexts. Cloud-based servers allow facilitator and student use whenever and from wherever Internet access is available. Cloud-based servers are thought to be risky in terms of data security and because of the need for Internet access in developing countries. A third option of a hybrid hosting system (https://seertechsolutions.com/our-services/hosting-options) could be used to counter the conflicting concerns of accessibility and perhaps data security but do not address the need for technical support and reliable local infrastructure. ELearning is likely to involve users needing to access wider material from the internet or course content from a remote location (Romiszowski 2013; Pocatilu et al. 2009). MLearning is an extension of eLearning, where increasing access to mobile computing and wireless technologies allows users to access their learning anywhere (Garcia-Cabot et al. 2015). These drivers, the reduced need for specialist technical staff, and the potentially greater data management capabilities in the cloud lead the design decisions toward a cloud-based server delivery mechanism.

## 4.2 Element Design

A critical step in Cycle 3 of the ADR approach is to determine which course elements can be meaningfully employed given the developing country context. ICT has already shown its benefits within developed countries, including web 1.0 and early 2.0 interactions, and the use of rich interactive media content, social networks and collaborative spaces. These technology and eLearning trends from developed countries can be used to create course elements that are engaging and effective (Clark and Mayer 2011). In this research, the curriculum is already developed, thus the element design focuses on delivering the established learning outcomes through the eLearning platform, reflecting the context of the target developing country.

There is a real need for course designers to recognise the differences in the andragogies of most developed and developing countries. Andersson and Grönlund (2009, p 1) have stated: "the active, participative student that is required for interactive learning is […] very rare in countries where the tradition is to teach in a more didactic manner"; therefore, element design may need to reflect a more classical form of instruction whilst still being a self-motivating experience. The simplest element design would be to digitise the course into a set of PowerPoint slideshows, build a reading list on a compact disk, and provide it to all the course participants. This simple approach does not exploit the potential of the Internet medium, but can work when the learners have already had experiences with self-paced intrinsically or extrinsically motivated learning.

As explained earlier, developing country tertiary education providers have been early adopters of eLearning. This early adoption could be explained by students, facilitators and managers already being part of an educational institution, a younger more technically adept demographic of users, or a higher focus on the progression and advancement of learning. Providers of vocational and workplace training in developing countries have yet to exploit current popular technologies to provide the same outcomes as their tertiary counterparts. For this reason, element design will focus on the attitudes of individuals and ensuring that the eLearning platform and elements are familiar, user friendly and engaging, for the students and the facilitators. Course designers should be exploiting the users' use of social media and popular Internet technologies to develop edutainment, serious gaming and online communities of practice (Romiszowski 2013).

### 4.3 Content

The main part of Cycle 3 of the ADR approach we have taken is content development. ELearning is an ideal platform for rich content to be delivered to students. The curriculum in this research has been developed by training developers in conjunction with subject matter experts. An initial hurdle is what type of electronic presentation of the content will best suit the students and if it is feasible to be developed. Developing countries may have limited access to high-end content development facilities as well as low comfort levels with ICT and usability issues with new systems (Ssekakubo et al. 2011). Gemmell et al. (2015) have looked into how international students overcame cultural differences in learning styles and discovered that a range of teaching methods including interactive learning objects, simulations, links to external materials, videos, audio files, and digitised textbooks were effective. This form of multimodal content aligns with the principles and processes of learning, known as: Multiple Channels—the ability to learn using visual, auditory and kinaesthetic (VAK) means; Limited Capacity—the amount of learning a person can do in each channel at one time; and, Active Processing—the active cognitive participation in learning about the content not just remembering facts (Clark and Mayer 2011). Multimodal means that the content should be delivered in a multitude of ways. Content should be combined into comprehensibly-sized elements, and it should be engaging and interactive. ELearning is inherently good at providing design opportunities for multimodal delivery of content, as it can take advantage of the digital device to provide VAK experiences. An eLearning platform can deliver this in a variety of ways, which could include a mix of the items within Table 4.3:

| Visual | Auditory | Kinaesthetic |
|---|---|---|
| Text on screen | Video | Activities |
| Links to other readings | Podcast | Scenarios |
| Video | Tele/video conferences | Problems solving |
| Info-graphics | | |

*Table 4.3. VAK resources in eLearning (based on Clark and Mayer 2011)*

In its traditional sense, kinaesthetic learning meant working with one's hands to learn, but in the case of eLearning it includes active participation in scenarios, problem solving, and simulation to provide the hands on learning (Christen 2009). The multimodal approach to learning including VAK and Problem

Based Learning has been found to also be beneficial to students in developing countries (Alkhasawneh, Mrayyan, Docherty, Alashram, and Yousef 2008), but it is important to contextualise the use of these modes to the host country. In order to present the content electronically, course designers must know what tools are available to produce and deliver material that best suits the needs of developing countries, such as:

**Video Creation and Hosting.** Jung and Lee (2015) explain that the use of video (both creation and viewing) through online hosting sites, such as YouTube, can be used practically to assist in cross cultural education, specifically in its use of making lectures available for later use. With the availability of video editing software they can also go another step further in providing effects such as info-graphics, kinetic typography and closed captioning to assist in providing additional contextual multimodal delivery of content. Jung and Lee (2015) also state that not enough research has been given to the implications of the differences in age and gender of the presenter used on the videos and it would also be expected that the use of language, accent and cultural background could be contributing factors to the acceptance of videos as effective learning content.

**Social Media.** First time users of eLearning can feel uneasy and anxious, similar to the idea of culture shock in international students studying abroad. This could be due to the loss of familiar educational practices, symbology, language and social intercourse (Xing and Spencer 2008), which can lead to the resentment toward the host system. This culture shock should be considered when designing content and is where the use of social media can help. If students are encouraged and provided with opportunities to connect with other students from the same cultural background they can find common ground within the course and prevent the feeling of isolation or uneasiness (Xing and Spencer 2008). The networks built through these social and collaborative activities help in the motivation and ongoing learning (Gemmell et al. 2015).

**Collaborative Workspaces.** The use of forums, cloud-based documents and other collaborative workspaces can also help to create connections between students. Gemmell et al. (2015) promoted the use of forums as an interactive community of inquiry, which provided critical interactions between students to increase students' success. Forums have been found to provide motivation for learner participation in the course, as well as reinforced learning through peer-review of posts, which also provides the facilitator with an overwatch and intervention point if required to bring students back on task where required (Organero and Kloos 2007).

**Problem Solving Activities.** The use of problem solving activities can reduce cultural barriers in learning. Problems can be introduced in a gradual progression for those that are new to use of ICT and eLearning, beginning with the use of structured materials such as embedded multimedia and hyperlinks, and progressing onto the use of web searches to search the wider internet. This scaffolded approach will help students to build skills in their ICT and Internet technologies use to further develop their problem solving skills (Xing and Spencer 2008).

# 5   ICT Artefact Conceptual Prototype Design

In keeping with the ADR approach of *Reflection and Learning* within cycles, this section presents the principles-based outline design developed in this research for the ICT artefact.

**Platform Selection**. The criteria for the EGMC are: the requirement to deliver the training to over 1000 personnel (to be confirmed), students will be located throughout Bangladesh, central course management and facilitation from the BPATC, the course should be self-paced, and the course should require minimal effort from the facilitator.

The platform that best suits the above criteria is the MOOC, which provides the benefits of customisation, usability, range of content, easy enrolment of large numbers of students and less facilitator intervention required. The disadvantages of using MOOC will be addressed as part of the content design, to ensure that it is engaging as well as making sure that students feel that they are part of a community of learning and are not isolated.

**Platform Delivery Selection**. Cloud computing provides a viable option with benefits of lower set-up costs, connection reliability, data security, server reliability, and less technical assistance required to maintain equipment. This will help to increase the positive attitudes toward ICT use as well as higher user satisfaction levels. The risk to data security and the need for Internet access are much less problematic than the need for local technical staff and reliable local infrastructure in the delivery of the eLearning platform in this case.

**Content Selection**. As mentioned, a multimodal approach to content delivery is the best method of course design and therefore the element design will use multimedia rich video, social media, collaborative workspaces, and problem solving activities. Importantly, the content will be contextualised for the target country, Bangladesh. Each topic will need to be delivered in a variety of ways within the ICT artefact to give the participants a range of experiences so they can compare and provide feedback on the design that best suits their needs. The elements will also be designed to unburden the facilitator, by giving the students the resources, social networks, and links to find the information for themselves. Collaborative activities will also involve peer review to assist in group moderation leaving the facilitator to observe and react where necessary as well as some marking of work.

**Course Design**. With the selection of the platform and online hosting, the initial course development can begin. During this initial setup the platform will be built to provide a container for the supporting course elements. The instructional design analysis of the current course content will be conducted on the face-to-face EGMC lesson packages and textbook to understand the required level of content and how it would best be chunked and delivered in an eLearning environment.

# 6  Conclusion

This paper provides the *Reflection and Learning* on the first three cycles of the ADR, specifically what has been explored during the initial planning and design of the ICT artefact. ELearning is now a viable alternative to mainstream education and an equivalent and often preferred method of learning. We asked, can the challenges of eLearning in developing countries be overcome with course design, using current eLearning trends? We found that developing countries are following their developed counterparts in the adoption of eLearning. With some recent improvements in the ICT infrastructure, attitudes toward technology, and the falling prices for consumer computing goods, the digital divide may be on the mend (Wijetunga 2014). The notion of digital citizens and digital natives has meant that people are finding new ways to consume information (Romiszowski 2013; Imran et al., 2013b). Creators of eLearning everywhere must now exploit how citizens currently use ICT, being digital entertainment, gaming and social media, as well as problem solving through the use of search engines to discover answers. We demonstrated that the trends can be implemented through the design of the course its elements in the context of the user, using multimodal forms of content delivery to create eLearning products. Content and element design must consist of contextualised multimodal VAK content delivery to ensure the content is understood by as many users as possible (Clark and Mayer 2011), and delivered in a form that users are familiar and comfortable with (Xing and Spencer 2008). Element design paired with social and collaborative interactions means the user will be more accepting of the system and part of the community of learning (Gemmell et al. 2015).

The use of the ADR process has meant that the development of an ICT artefact, using the research from the *Problem Formulation* stage, will provide evidence as to whether our selected course and element design can overcome challenges of eLearning in developing countries. The early stages of the ADR have allowed us to understand which approaches may be more effective and has given us the opportunity to develop a tool to use, assess, redesign and reuse. Our research has moved into the later stages of the ADR process, where the ICT artefact is being refined, and will be deployed to users in Bangladesh as part of the *Intervention* stage. The ADR stages will be concluded with the *Formalisation of Learning* stage, which will provide the findings of the ICT artefact *Evaluation*, as well as the *Reflection and Learning* gathered from all cycles, to be presented in the phase two paper.

# Copyright